# Photoinduced anisotropic lattice dynamic response and domain formation in thermoelectric SnSe

Wei Wang[1,#], Lijun Wu[1,#], Junjie Li[1], Niraj Aryal[1], Xilian Jin[1,2], Yu Liu[1,#], Mikhail Fedurin[3], Marcus Babzien[3], Rotem Kupfer[3], Mark Palmer[3], Cedomir Petrovic[1], Weiguo Yin[1], Mark P. M. Dean[1], Ian Robinson[1,4], Jing Tao[1], Yimei Zhu[1*]

[1.] Condensed Matter Physics and Materials Science Division, Brookhaven National Laboratory, Upton, NY 11973

[2.] State Key Laboratory of Superhard Materials, College of Physics, Jilin University, Changchun 130012, China

[3.] Accelerator Science & Technology Initiative, Accelerator Test Facility, Brookhaven National Laboratory, Upton, NY 11973

[4.] London Centre for Nanotechnology, University College, London WC1E 6BT, UK

[#] These authors contributed equally: Wei Wang, Lijun Wu

[#] Current address: Los Alamos National Laboratory, MS K764, Los Alamos, NM 87545, USA

*Corresponding author: zhu@bnl.gov

## Abstract

Identifying and understanding the mechanisms behind strong phonon-phonon scattering in condensed matter systems is critical to maximizing the efficiency of thermoelectric devices. To date, the leading method to address this has been to meticulously survey the full phonon dispersion of the material in order to isolate modes with anomalously large linewidth and temperature-dependence. Here we combine quantitative MeV ultrafast electron diffraction (UED) analysis with Monte Carlo based dynamic diffraction simulation and first-principles calculations to directly unveil the soft, anharmonic lattice distortions of model thermoelectric material SnSe. A small single-crystal sample is photoexcited with ultrafast optical pulses and the soft, anharmonic lattice distortions are isolated using MeV-UED as those associated with long relaxation time and large displacements. We reveal that these modes have interlayer shear strain character, induced mainly by **c**-axis atomic displacements, resulting in domain formation in the transient state. These findings provide an innovative approach to identify mechanisms for ultralow and anisotropic thermal conductivity and a promising route to optimizing thermoelectric devices.

## Introduction

Thermoelectric materials provide a platform to study phonon-phonon interactions, which is a central topic in condensed matter physics[1–4]. The dimensionless thermoelectric figure of merit $(ZT = \frac{S^2\sigma}{\kappa}T)$ implies that good thermoelectrics require low thermal conductivity, $\kappa$, and high electrical conductivity, $\sigma$, along with a large Seebeck coefficient $S$[5,6]. To optimize the thermal transport in materials, it is crucial to understand the correlations among the phonon modes and how they impact the thermodynamics. A good example is thermoelectric tin selenide (SnSe), which has recently attracted tremendous attention due to its ultrahigh $ZT$ of ~ 2.6 at 923 K[7,8]. SnSe is a semiconductor with an indirect electron band gap ~ 0.86 eV[9–12], therefore, phonons are the dominant heat carrier[13–15]. Its high $ZT$ value has been associated with soft, anharmonic distortions off the lattice, which cause strong phonon-phonon scattering [16–18]. The SnSe crystal structure features Sn-Se zigzag-chains along the **b** axis and armchair-chains along the **c** axis[19]. These **b-c** planes are stacked along the **a** axis to form an orthorhombic unit cell with strong anisotropy in electronic conductivity, phonon dispersion relation, and optical properties[9,16,20,21]. Around 800 K, SnSe undergoes a second-order phase transition between two different symmetry orthorhombic phases from *Pmna* (#62) phase to *Cmcm* (#63) phase, which is associated with the condensation of a soft optical phonon mode with an $A_g$ symmetry[22,23].

Here we report the use of an ultrafast pump-probe approach to directly unveil the soft, anharmonic lattice distortions involved in the strong phonon-phonon scattering in SnSe. These are identified by exploiting the fact that these soft, anharmonic distortions are associated with long recovery timescales after photoexcitation. The ensuing motions are tracked by MeV electron diffraction through its access to large regions of momentum space[24,25]. We demonstrate a photoinduced domain formation and evolution in the **b-c** plane in the transient state at 90 K. Combining quantitative ultrafast diffraction analysis and first-principles calculations, we find that a soft phonon mode, transverse optical (TO) $A_g$ phonon mode at the Brillouin zone center in the *Pnma* phase, is highly excited[23]. The strong anharmonicity of the phonon mode induces atomic displacements from their equilibrium positions mainly along the **c** axis. The lattice distortion corresponding to the $A_g$ phonon modes introduces different strain effects to the system. Their competition gives rise to a shear strain between the layers, which significantly reduces the correlation length along the **c** axis via lattice distortion, yielding in-plane domain nucleation. The domain boundaries likely introduce additional diverse phonon scatterings to reduce thermal conductivity. Furthermore, the contributions from these phonon modes are sensitive to the sample temperature as the strain generation and the domain formation in SnSe are suppressed at 300 K.

## Results and Discussion

**Bragg peak intensity evolution.** Figure 1(a) shows a schematic of the MeV-UED experimental setup, where we record diffraction patterns along [100] zone from a single-crystal SnSe as a function of time delay during the pump-probe process. The normalized diffraction intensities $I/I_0$ ($I_0$ being the peak intensity before time-zero) of (0$k$0) and (00$l$) Bragg peaks as a function of time delay are shown in Fig. 1(b). Upon photoexcitation, the intensities of all the Bragg peaks were found to decrease. The intensity of {040} reflections, averaged over its equivalent reflections, starts to bounce back around 50 ps, while the intensities of other Bragg peaks, such as, {020}, {002}, {004}, remain low. In Fig. 1(c) and (d) we plot the intensities of the {002} and {004} reflections as a function of random thermal vibration (Debye-Waller factors) and atomic displacement, respectively, based on structure factor calculations (for details see SI). The calculation illustrates that with the increase of the thermal vibration $\langle u^2 \rangle$ the diffraction intensity decreases more rapidly for high-index q {004} than for low-index q {002}. For atomic displacement, however, the intensity deceases with the increase of the displacement $\Delta z$ of Sn for {002} but increases for {004}. If there only exist the changes of thermal vibrations, as described by thermal mean square displacement $\langle u^2 \rangle$, during the relaxation process after photoexcitation, the normalized intensity change ($\Delta I/I_0$) of {040} and {004} reflections would be larger than that of {020} and {002}, which is inconsistent with our experiments, as shown in Fig. 1(b). According to these calculations, specific atomic displacements should be triggered by the incident photons. During the first 50 ps after laser pumping, the relative peak intensity can be fitted with a single exponential decay function with a time constant of ~18.3±3.1 ps. The observed timescales support the idea that the atomic displacements correspond to the low energy-cost lattice distortion through electron-phonon interaction and phonon-phonon scattering (See Supplementary Fig. 1). The slow thermalization process is mainly due to the low thermal conductivity in SnSe. Therefore, a combination of thermal vibrations and atomic displacement may be responsible for the observed intensity variations with a rather slow time scale compared with other systems[26,27]. Additional intensity measurements for the Bragg peaks can be found in Supplementary Fig. 2.

**Asymmetrical intensity distribution of Bragg peaks.** The full effect of the phonon contributions to the electron diffraction is also encoded in the diffuse scattering surrounding the Bragg peaks. To monitor the subtle changes in diffraction intensities upon photoexcitation, we constructed experimental intensity difference maps, $\Delta I$ ($\mathbf{q}$, $t$) = $I$ ($\mathbf{q}$, $t$) − $I$ ($\mathbf{q}$, $t \leq 0$), where $t \leq 0$ represents the times before the pump arrives and $\mathbf{q}$ represents the in-plane momentum transfer. Figure 2(a) shows a typical difference map of the diffraction patterns for the time-delay of $t = 1.7$ ps. The figure shows the Brillouin zones of the (020) Bragg peak of SnSe in the *Pnma* phase (the full diffraction pattern is shown in Supplementary Fig. 3). We observe an angular dependence of the intensity distribution around the Γ points with elongation along the Γ-Z direction. The line profile of the Bragg peaks in the intensity difference map along Γ-Z and Γ-Y directions are shown in Fig. 2(b). Away from the Γ point center, the intensity sharply rises along the Γ-Z direction, indicative of a drastic intensity distribution change after photoexcitation. In contrast, along the Γ-Y direction there is little change. The asymmetrical intensity distributions around the Bragg peaks were only observed during the first ∼ 20 ps at 90 K. After that the intensity distribution in the difference maps becomes symmetrical again and decreases equally in all directions. Full details are provided in Supplementary Fig. 4, showing that the asymmetrical intensity distribution of the Bragg peaks changes with the time delay.

**<u>The origin of the asymmetrical intensity distribution.</u>** The asymmetrical intensity distribution in the difference maps of the Bragg peaks in time domain is unexpected and cannot be explained by short-wavelength phonons since there is no observable diffuse scattering in the rest of the Brillouin zone of the UED patterns, further away from the Bragg peaks. Inelastic scattering arising from specific phonon modes as the cause can also be excluded [28].

Instead, the variation of reflection peak width can be related to the change of the correlation length ($\xi$) of a mosaic domain structure in real space. The equations 2 in Supplementary note 5 describe the scattering intensity for a finite-sized domain, i.e., correlation length, with dimension of $\xi_a$, $\xi_b$ and $\xi_c$ along **a**, **b** and **c** direction, which may also be applied to the situation where there are a large number of domains scattered incoherently by incident electrons[29]. Supplementary Fig. 4 shows the shapes and peak width of the reflection for different domain sizes along the **b** and **c** directions when viewed along the **a** direction. For domains with $\xi_b = \xi_c$, the reflection intensity profile is symmetrical. Domains with $\xi_b \neq \xi_c$ lead to an asymmetrical peak broadening, e.g., elongated along the Γ-Z direction if $\xi_b > \xi_c$. The resulting difference map (Supplementary Fig. 5(c)) captures exactly what we observed in the experiment. We therefore attribute the observed asymmetrical intensity distribution of the Bragg peaks to the temporal formation of domains with specific aspect ratios in response to the photoexcitation.

Our results are consistent with the formation of domains that have various shapes and sizes, arising from the photoinduced lattice distortion in SnSe. Considering the scattering among these photoinduced domains with small lattice distortions, we perform diffraction simulations based on the fact that in our UED measurements the Bragg peak positions remain unchanged before and after photoexcitation and there is no detectable lattice parameters or volume change in the system. We use Monte Carlo and dynamic multislice method by constructing a large supercell consisting of $n_b \times n_c \times 1$ ($n_b$, $n_c \geq 64$) SnSe unit-cells along $\mathbf{b}$, $\mathbf{c}$, $\mathbf{a}$ direction with sampling points of $4096 \times 4096$ for each slice with thickness of 0.29 nm. The supercell is composed of domains with averaged domain size of $m \times n \times 1$ unit-cells, where m and n determine the domain dimension along the $\mathbf{b}$ and $\mathbf{c}$ directions. We consider the lattices inside each domain are rigid or uniformly distorted based on the corner positions of the domains that deviate from their equilibrium state (Fig. 2(f)). The small deviation ($\leq 1\%$) of the domain corner follows a statistically random distribution both in space and in time, accounting for the domains with different distortions in the probing area and the frozen supercells with various distorted domains at a particular time, i.e., at each single-shot acquisition, respectively (see Method section). This approach is consistent with our experimental setup where a UED pattern is recorded at certain time delay through an accumulation of many single-shots. The electrostatic potentials of the supercell were then used to calculate the diffraction patterns based on the multislice method, as shown in Fig. 2(g) and Supplementary Fig. 6. We refer the details on the Monte Carlo and multislice simulations and domain configurations to Supplementary Note 6.

Figure 2(c) shows simulated intensity distribution in a difference map, where there is a reduction in domain dimension by a factor of four along the $\mathbf{c}$ direction and no change along the $\mathbf{b}$ direction after photoexcitation (Fig. 2(e)). The intensity features in the simulated diffraction pattern agree well with the experimental observations, as shown in the corresponding line profiles along the Γ-Z and Γ-Y directions in Fig. 2(b). Figure 2(e) depicts the domain formation, evolution, and annihilation dynamics based on the Bragg peaks analysis from UED patterns at different time delays (for details, see Supplementary Fig. 4). Figures 2(f, g) show the lattice distortion (exaggerated for clarity) of the domain and the corresponding projected electric potential. For comparison, the potential for undistorted SnSe crystal is shown in Fig. 2(h). Upon photoexcitation, the initial crystal structure of SnSe is fragmented into small mosaic domains by mainly reducing the domain size along the **c** direction. For instance, the domain size along the **c** axis is reduced by factor of two at around 1 ps and by a factor of four at around 8 ps, at which the shoulder of the peak intensity reaches the maximum. Subsequently, the small domains annihilate, and eventually the system enters a metastable state with comparable domain sizes along the **b** and **c** directions after 20 ps.

We note that the elongation direction of the Bragg peaks does not always point to the Γ-Z direction (Supplementary Fig. 4), but varies in direction at different time delays over a small angular range. This implies that the photoinduced domains may rotate slightly around the [100] direction (Γ-Z direction in reciprocal space). As described in Supplementary note 7, the effect of domain rotation can be simulated using the same approach by considering both distortion and different rotations of the domains in the supercell (Supplementary Fig. 8(d)). Similar to the Γ-Z oriented unrotated domains, here the averaged SnSe lattice parameters and orientation remain unchanged, being consistent with the unchanged Bragg positions observed in UED experiments. When the domain rotation away from the [100] direction increases, the lattice misfit strain and interfacial energy at the domain boundaries increases. This is likely the reason that the elongation direction of the reflections in UED experiments exhibits a small angular distribution around Γ-Z direction.

**Temperature-dependent behavior.** The intensity variations of Bragg peaks at 300 K show a similar trend compared with those observed at 90 K (Fig. 3(a)), implying that atomic displacements also take place at 300 K during the photoexcitation process. However, the Bragg peak intensities drop faster at 300 K (time constant $\tau_{300K} \sim 1.7$ ps), comparing to that at 90 K ($\tau_{90K} \sim 20$ ps) after the laser pump (Fig. 3(b) and Supplementary Fig. 9). The thermal diffuse scattering (TDS) intensity in Fig. S1 the saturated TDS intensity at 300 K is smaller than that at 90 K, which demonstrates the photoinduced increase of lattice temperature at 300 K ($\Delta T_{lattice\_300K}$) is less than that at 90 K ($\Delta T_{lattice\_90K}$). The smaller lattice temperature variation is related to the larger heat capacity in SnSe at 300 K. Additionally, the TDS intensity at 300 K saturates at a relatively shorter time delay compared with that at 90 K, which could be associated with the small temperature jump ($\Delta T_{lattice\_300K}$) at 300 K. In terms of the faster response ($\tau_{300K} \sim 1.7$ ps) at 300 K, apart from the thermalization impact, atomic displacements following the phonon modes also play an important role in the intensity variation. F. Knoop *et al.* proposed "anharmonic force" associated with the degree of phonon anharmonicity, which governs the atomic displacements[30]. In our case, we infer that the anharmonic force of the phonon modes is larger at 300 K in SnSe, which would drive the atoms move faster. Importantly, the asymmetrical intensity distribution, not observed at 300 K (Fig. 3(c), and Supplementary Fig. 10), only exists for the photoexcitation process at low temperatures and likely not thermally driven. The absence of asymmetrical intensity distribution infers that there is no domain structure with a specific shape and size generated at and above room temperature.

**Possible driving force for the domain formation.** To understand the origin of the domain formation, we performed the DFT calculations (Supplementary note 11). After tens to hundreds of femtoseconds, the '*hot*' electrons relax to thermal distribution described by electron temperature $T_e$, which is much higher than sample's base temperature. Due to their larger predicted electron-phonon coupling the higher-energy optical phonons are more likely to be excited than the acoustic modes, yielding a nonthermal distribution of phonons[31–34]. Figure 4(a) shows the calculated partial density of states for $T_e = 0$ K (light color) and $T_e = 6000$ K (dark color), where the 1.55 eV photons excite the electrons just below the Fermi surface to the conduction band. The changes in the electron density of state manifest the simultaneous hole doping into the valence bands and electron doping into the conduction bands.

The DFT calculations predicted the anharmonic lattice distortions from the ground state to a few finite-$T_e$ structures, are shown in Fig. 4(b). They are characterized by a decrease in the intralayer lattice distortion but an increase in the interlayer lattice distortion, instigating a shear motion along the **c** axis. Since the structural relaxation with a finite $T_e$ does not change the lattice symmetry, the $T_e$ induced atomic displacements can be expressed as a superposition of all the four $A_g$ modes referred to $A_g$-5, $A_g$-12, $A_g$-19, $A_g$-23 based on their energy ranking, as shown in Figs. 4(c) and 4(d). The calculated results indicate that atomic displacements are mainly governed by phonon modes $A_g$-5 and $A_g$-23. While the soft phonon mode $A_g$-5 represents the thermal phase transition, the shear strain is tied to the highest-energy $A_g$-23 mode. The projected atomic distortions induced by these two strongly anharmonic phonon modes are in the opposite direction in the **b-c** plane. Hence the atomic displacements following these two competitive phonon modes introduce a shear strain along the **c** direction as shown in Fig. 4(b), and the similar photoinduced strain formation has been reported in [35]. The extra shear strain can be relieved by inducing lattice deformation in **b-c** plane and reducing the corresponding correlation length, which results in domain formation in the **b-c** plane[36]. Our domain structure model for the diffraction simulation is based on this domain formation mechanism. Due to the introduced shear strain, the atomic position, i.e., corners in each block, can be deviated from the original positions to new positions. And the nearby atoms distort following the displacements of the corner atoms. The atom deviation is random, which reduces the correlation length, i.e., domain formation. Since the atomic displacements are mainly along **c** direction in SnSe during the photoexcitation, the correlation length in **c** axis is highly reduced, that is, the domain size along **c** axis is smaller.

In the domain structure model, the domain boundary is defined by the corners of each block, which are outlined by red and green lines in Supplementary Fig. 6a and 6d. Additionally, the dense domain boundaries contribute to phonon scatterings, including Rayleigh scattering, Umklapp scattering and displacement scattering, and suppress the lattice thermal conductivity. Upon subsequent release of the shear strain, the opposite pathway of atomic displacements leads to a recovery of the **c**-axis domain size, or correlation length. The electron diffraction calculations incorporating domain structures and atomic displacements are shown in Supplementary Figs. 6-8. The results illustrate that the domain size dominates the asymmetrical intensity distribution around the Bragg peaks. Based on our experimental observation, the photoinduced domain size starts to increase after $t_2 \sim 8$ ps in Fig. 2(d). At longer time delay, $t > 20$ ps, the asymmetrical intensity distribution disappears, suggesting the correlation length difference in the **b** and **c** direction vanishes. Then the atomic displacement following $A_g$-5 phonon mode and the Debye-Waller factor dominant in the intensity variation. The calculated diffraction intensities induced by the atomic displacement corresponding to the $A_g$-5 phonon mode are shown in Supplementary note 12. Regarding the UED observation at 300 K and the DFT calculation result, the absence of asymmetrical intensity distribution infers that the light-induced atomic displacement is likely dominated by the $A_g$-5 mode at higher temperatures, which is related to the structure phase transition. The absence of the competition among these phonon modes suppresses the shear strain formation and the domain generation at 300 K.

In SnSe, the Sn-Se bonds along **a**, **b** and **c** axes are anisotropic and anharmonic, which leads to a strong phonon scattering and low thermal conductivity[7,8,16]. Among these bonds, it has been pointed out that **c**-axis Sn-Se bond strength is notably weak, which enables the atoms to move preferentially along **c** axis[16,23]. Additionally, there is a negative thermal expansion (NTE) in this direction, which could be related to the electronic instability and its coupling to the anharmonic vibration of atoms along the **c** axis[22,37]. Additionally, the lattice distortion involved in the NTE is strongly coupled to the anharmonic $A_g$ optical phonon mode[17,37]. During the photoexcitation process, the atomic displacements corresponding to the selected $A_g$-5 and $A_g$-23 optical phonon modes have been realized via electron-phonon coupling at 90 K. Due to the far-from-equilibrium state, the interlayer incompatibility from the lattice displacements introduces a shear strain, reduces correlation length and generates domain structures. The photoinduced strain evolution is incorporated into tens of ps time scale, which is comparable with the timescale observed in [35]. It is well known that phonon, as the heat carrier, plays a key role in the thermal transport. In laser pumped SnSe, with the induced domain structures discovered here, the enhanced phonon scatterings at the domain boundaries could shorten the mean free path of phonon propagation in the crystal. In this case, the involved phonons range could be from low to medium frequency[38], and further reduce thermal conductivity of SnSe, leading to an improved thermoelectric performance.

Bring all pieces together as shown in Fig. 5, we can understand the emergence of domain formation after photoexcitation as following. The electrons are excited on the short (tens of femtoseconds) time scale and the corresponding electron temperature could reach thousands of Kelvin. Then, the energy of the electrons is transferred to the lattice through electron-phonon coupling. Due to the strong anharmonicity of phonon modes in SnSe, atomic displacements coupled to anharmonic phonon modes are realized. Based on our experimental observations and DFT calculations, the atomic displacement along **c** axis driven by competing $A_g$ phonon modes, dominates in the first 20 ps after photoexcitation at 90 K. The relative Sn-Se atomic displacement introduces a shear strain in the interlayers, which causes more lattice deformations and reduces the correlation length mainly along the **c** direction. Namely, domains with a specific dimension are generated in the transient state. Additionally, the domain may slightly rotate in the **b-c** plane to accommodate strain, which impacts on the domain corner positions and the overall lattice distortion in the domain structure. The release of strain enables the atoms to move back to the original positions following other $A_g$ phonon modes, which gradually increases the domain size along the **c** direction. The absence of domain formation at 300 K infers that the phase-transition related $A_g$-5 phonon mode dominants after photoexcitation, which significantly reduces the shear strain in the **b-c** plane. Summarizing our findings, the relatively long-time scale, i.e., tens of ps, observed in our UED experiment is a combination result from the low thermal conductivity in SnSe, lattice displacements induced by the anharmonic phonon modes and the shear strain. Given the close connection to anharmonicity, a non-linear laser-fluence dependence would be expected, which will be tested in future experiments.

In conclusion, using MeV-UED we have observed unexpected anharmonic lattice distortion, which launches a transient domain formation in single-crystal SnSe following photoexcitation. We reveal that this domain formation in the non-equilibrium state is due to an interlayer shear strain, induced by atomic displacements along the **c** axis. DFT calculation indicates that the observed atomic displacements correspond to TO $A_g$ phonon modes with strong anharmonicity and phonon-phonon scattering in SnSe. Additionally, the transient state of domain structure state only survives for ~20 ps at 90 K. The domains are created by incident photons, and then start to merge and grow during a strain release process. Our observations demonstrate that ultrafast optical pulses are capable of not only manipulating transient and metastable material properties but also creating lattice structures that are not known to exist in the thermal equilibrium state. We expect the domains and their boundaries we observed can be important sources for phonon scattering in reducing thermal conductivity and further improving thermoelectric performance in SnSe.

## Methods

**MeV UED.** The MeV-UED experimental setup at BNL with 3 MeV electron pulses of < 180 fs duration transmitted through the sample at the normal incidence, similar to that reported previously[39,40]. A laser pulse with a duration of 180 fs at 800 nm wavelength and pump fluence of 1.0 mJ cm$^{-1}$ was used in the experiment which is sufficient to produce a photoinduced lattice dynamic process but prevent photoinduced sample damage. Overall temporal resolution considering RF synchronization jittering (100-150 fs), pump laser duration (180 fs), probe duration (<180 fs), horizontal probe beam size (100-300 um) and angle between pump and probe (15 degree), we have a temporal resolution between 250 to 350 fs. The SnSe single crystal with layered orthorhombic crystal structure was exfoliated along the (100) plane and suspended on a 3 mm Mo grid. Measurements were performed at liquid nitrogen temperature (~ 90 K) and room temperature. Since the beam direction in the UED experiment is parallel to the **a** axis, our probe is more sensitive to the lattice distortion in the **b-c** plane. A total of 54 Bragg peaks are captured and analyzed. A UED pattern is recorded by accumulating 72 single-shots. We repeat each measurement 10 times and the results were averaged to ensure the data is statistically meaningful. The intensities of equivalent $(hkl)$ and $(\bar{h}k\bar{l})$ reflections are further averaged as $I_{\{hkl\}}$ to improve the signal noise ratio.

**UED simulations.** To understand the origin of the intensity elongation in the difference maps, we developed computer codes to calculate UED patterns based on Monte Carlo and dynamical multislice methods. To account for the photoinduced strain and domain formation we built a large supercell which consists of many domains. We consider the lattices inside the domain are rigid and distorted along with the corner positions of the domains when they are out-of-equilibrium (Fig. 2(f), Supplementary Fig. 6 and Fig. 8). The deviation of the domain corner from their equilibrium positions exhibits statistical randomness following a Gaussian distribution with its mean (or equilibrium position) at the center of the bell-shaped curve. The supercell is used to calculate the UED pattern with the multislice program developed in-house[41]. More than 20 of such a UED pattern, each corresponding to a different supercell configuration with a random corner position, were then calculated and averaged to account for the multi-shots UED experiment when a diffraction pattern is recorded.

**First-principles calculations.** First principles simulations were performed by density functional theory in generalized gradient approximation implemented in Vienna Ab-initio Simulation Package. The projected augmented wave method was adopted for the potentials of Sn (5s25p2) and Se (4s24p4) atoms. The finite-displacement method was used to compute the phonon dispersion as implemented in Phonopy code. (See details in Supplementary Note 11)

## Data availability
All datasets generated during and/or analyzed during the current study are available from the corresponding author on reasonable request.

## Code availability
The code used for the dynamic diffraction simulation is available from L.W. on reasonable request.

## Acknowledgements
The authors gratefully acknowledge I. A. Zaliznyak for useful discussion. This research was supported by US Department of Energy (DOE), Office of Science, Office of Basic Energy Sciences (BES), Materials Sciences and Engineering Division under Contract No. DE-SC0012704. The BNL MeV-UED at the Accelerator Test Facility (ATF) is the DOE Office of Science User Facility for the Accelerator Stewardship Program. This work was also supported by the resources of Center for Function Nanomaterials (CFN) at BNL.

## Competing interests
The authors declare no competing interests.

## Author contributions
Y.Z., W.W., and J.L. conceived the idea of this study and planed the experiments and calculations. J.L. carried out the UED experiments and L.W. developed the method for Monte-Carlo-based dynamic diffraction simulations. M.F., M.B., R.K., and M.P. at the BNL-UED facility assisted the experiments. N.A., X.J., and W.Y performed first-principles calculations. Y.L. and C.P. synthesized the single crystal. W.W. prepared the UED samples. L.W. and W.W. performed data analysis of the UED experimental data and diffraction simulation. Y.Z., W.W., J.L., L.W., J.T., N.A., W.Y., I.R., and M.P.M.D. discussed and interpreted the data. W.W., L.W., and Y.Z., wrote the manuscript with inputs from all authors. W.W and L.W. contributed equally to this work.

# References

1. Sootsman, J. R., Chung, D. Y. & Kanatzidis, M. G. New and Old Concepts in Thermoelectric Materials. *Angew. Chem. Int. Ed.* **48**, 8616–8639 (2009).

2. Bell, L. E. Cooling, Heating, Generating Power, and Recovering Waste Heat with Thermoelectric Systems. *Science* **321**, 1457–1461 (2008).

3. Minnich, A. J., Dresselhaus, M. S., Ren, Z. F. & Chen, G. Bulk nanostructured thermoelectric materials: current research and future prospects. *Energy Environ. Sci.* **2**, 466 (2009).

4. Biswas, K. *et al.* High-performance bulk thermoelectrics with all-scale hierarchical architectures. *Nature* **489**, 414–418 (2012).

5. Gayner, C. & Kar, K. K. Recent advances in thermoelectric materials. *Prog. Mater. Sci.* **83**, 330–382 (2016).

6. Moshwan, R., Yang, L., Zou, J. & Chen, Z.-G. Eco-Friendly SnTe Thermoelectric Materials: Progress and Future Challenges. *Adv. Funct. Mater.* **27**, 1703278 (2017).

7. Zhao, L.-D. *et al.* Ultrahigh power factor and thermoelectric performance in hole-doped single-crystal SnSe. *Science* **351**, 141–144 (2016).

8. Zhao, L.-D. *et al.* Ultralow thermal conductivity and high thermoelectric figure of merit in SnSe crystals. *Nature* **508**, 373–377 (2014).

9. Pletikosić, I. *et al.* Band Structure of the IV-VI Black Phosphorus Analog and Thermoelectric SnSe. *Phys. Rev. Lett.* **120**, 156403 (2018).

10. Sirikumara, H. I. & Jayasekera, T. Tunable indirect-direct transition of few-layer SnSe via interface engineering. *J. Phys. Condens. Matter* **29**, 425501 (2017).

11. Hong, A. J. *et al.* Optimizing the thermoelectric performance of low-temperature SnSe compounds by electronic structure design. *J. Mater. Chem. A* **3**, 13365–13370 (2015).

12. Shi, G. & Kioupakis, E. Anisotropic Spin Transport and Strong Visible-Light Absorbance in Few-Layer SnSe and GeSe. *Nano Lett.* **15**, 6926–6931 (2015).

13. Ziman, J. M. *Electrons and Phonons: The Theory of Transport Phenomena in Solids*. (OUP Oxford, 2001).

14. Born, M. & Huang, K. *Dynamical Theory of Crystal Lattices*. (Clarendon Press, 1988).

15. Kang, J. S., Li, M., Wu, H., Nguyen, H. & Hu, Y. Experimental observation of high thermal conductivity in boron arsenide. *Science* **361**, 575–578 (2018).

16. Li, C. W. *et al.* Orbitally driven giant phonon anharmonicity in SnSe. *Nat. Phys* **11**, 1063–1069 (2015).

17. Liu, F. *et al.* Phonon anharmonicity in single-crystalline SnSe. *Phys. Rev. B* **98**, 224309 (2018).

18. Kang, J. S., Wu, H., Li, M. & Hu, Y. Intrinsic Low Thermal Conductivity and Phonon Renormalization Due to Strong Anharmonicity of Single-Crystal Tin Selenide. *Nano Lett.* **19**, 4941–4948 (2019).

19. Shi, L.-B. *et al.* Elastic behavior and intrinsic carrier mobility for monolayer SnS and SnSe: First-principles calculations. *Appl. Surf. Sci.* **492**, 435–448 (2019).

20. Zhang, C. *et al.* Anisotropic Nonlinear Optical Properties of a SnSe Flake and a Novel Perspective for the Application of All-Optical Switching. *Adv. Opt. Mater.* **7**, 1900631 (2019).

21. Yang, S. *et al.* Highly-anisotropic optical and electrical properties in layered SnSe. *Nano Res.* **11**, 554–564 (2018).

22. Chattopadhyay, T., Pannetier, J. & Von Schnering, H. G. Neutron diffraction study of the structural phase transition in SnS and SnSe. *J. Phys. Chem. Solids* **47**, 879–885 (1986).

23. Hong, J. & Delaire, O. Phase transition and anharmonicity in SnSe. *Mater. Today Phys.* **10**, 100093 (2019).

24. Li, R. *et al.* Experimental demonstration of high quality MeV ultrafast electron diffraction. *Rev. Sci. Instrum.* **80**, 083303 (2009).

25. Zhu, P. *et al.* Femtosecond time-resolved MeV electron diffraction. *New J. Phys.* **17**, 063004 (2015).

26. Eichberger, M. *et al.* Snapshots of cooperative atomic motions in the optical suppression of charge density waves. *Nature* **468**, 799–802 (2010).

27. Konstantinova, T. *et al.* Photoinduced dynamics of nematic order parameter in FeSe. *Phys. Rev. B* **99**, 180102 (2019).

28. Stern, M. J. *et al.* Mapping momentum-dependent electron-phonon coupling and nonequilibrium phonon dynamics with ultrafast electron diffuse scattering. *Phys. Rev. B* **97**, 165416 (2018).

29. Leoni, M. Domain size and domain-size distributions. in *International Tables for Crystallography Volume H: Powder Diffraction* 524–537 (2019).

30. Knoop, F., Purcell, T. A. R., Scheffler, M. & Carbogno, C. Anharmonicity measure for materials. *Phys. Rev. Mater.* **4**, 083809 (2020).

31. Beaurepaire, E., Merle, J.-C., Daunois, A. & Bigot, J.-Y. Ultrafast Spin Dynamics in Ferromagnetic Nickel. *Phys. Rev. Lett.* **76**, 4250–4253 (1996).

32. Waldecker, L., Bertoni, R., Ernstorfer, R. & Vorberger, J. Electron-Phonon Coupling and Energy Flow in a Simple Metal beyond the Two-Temperature Approximation. *Phys. Rev. X* **6**, 021003 (2016).

33. He, Z. & Millis, A. J. Photoinduced phase transitions in narrow-gap Mott insulators: The case of VO 2. *Phys. Rev. B* **93**, 115126 (2016).

34. Wall, S. *et al.* Ultrafast disordering of vanadium dimers in photoexcited VO $_2$. *Science* **362**, 572–576 (2018).

35. Zhang, B.-B. *et al.* Photoinduced coherent acoustic phonon dynamics inside Mott insulator Sr2IrO4 films observed by femtosecond X-ray pulses. *Appl. Phys. Lett.* **110**, 151904 (2017).

36. Kwak, B. S. *et al.* Strain relaxation by domain formation in epitaxial ferroelectric thin films. *Phys. Rev. Lett.* **68**, 3733–3736 (1992).

37. Bansal, D. *et al.* Phonon anharmonicity and negative thermal expansion in SnSe. *Phys. Rev. B* **94**, 054307 (2016).

38. Zhu, T. *et al.* Compromise and Synergy in High-Efficiency Thermoelectric Materials. *Adv. Mater.* **29**, 1605884 (2017).

39. Li, J. *et al.* Dichotomy in ultrafast atomic dynamics as direct evidence of polaron formation in manganites. *npj Quantum Materials* **1**, 1–7 (2016).

40. Li, J. *et al.* Probing the pathway of an ultrafast structural phase transition to illuminate the transition mechanism in Cu2S. *Appl. Phys. Lett.* **113**, 041904 (2018).

41. Wu, L. *et al.* Origin of Phonon Glass–Electron Crystal Behavior in Thermoelectric Layered Cobaltate. *Adv. Funct. Mater.* **23**, 5728–5736 (2013).

**Figures**

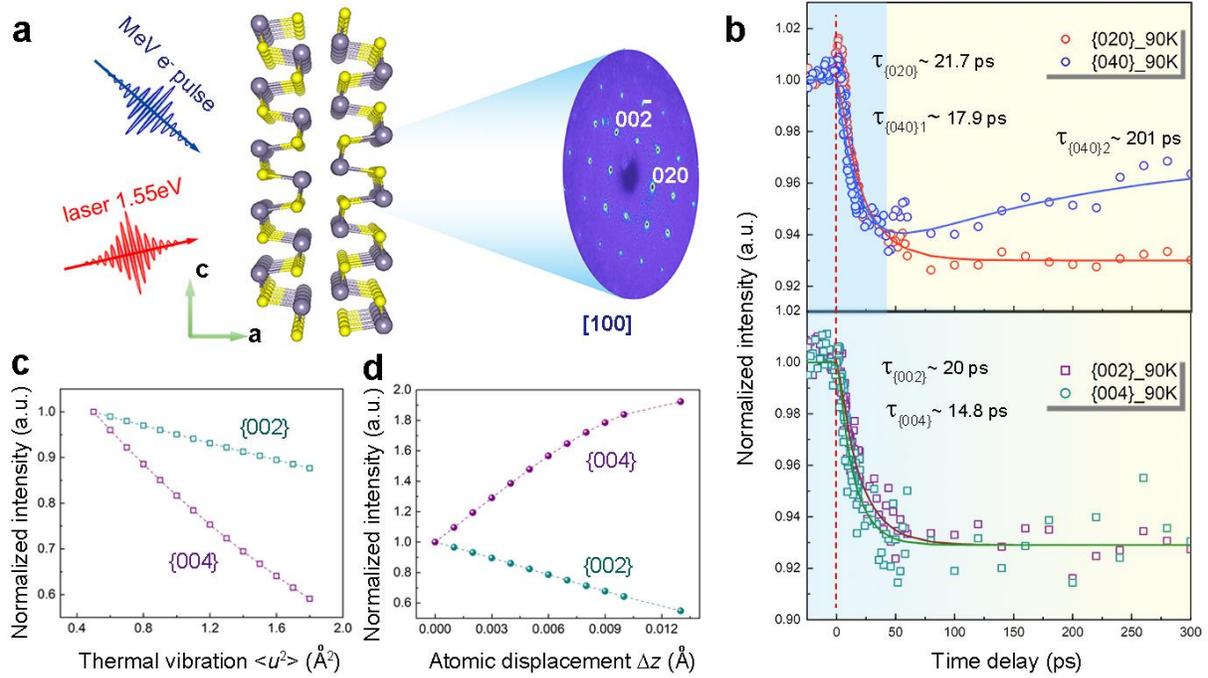

Fig. 1 **UED experimental setup and observations**. (a) Schematic of the ultrafast laser-pump electrons-probe setup to study photoinduced lattice dynamics in SnSe. The crystallographic axes are labeled. (b) Temporal evolution of the normalized diffraction intensity $I/I_0$ of Bragg peaks {020}, {040}, {002} and {004} at 90 K. The solid lines are fits of {020}, {002} and {004} using a single-exponential function; the solid line for the {040} peak intensity is fitted to a double-exponential function. The time constant for each plot is inserted. (c, d) Normalized intensity calculated based on structure factors as a function of thermal vibration $\langle u^2 \rangle$ (c) and atomic displacement $\Delta z$ for {002} and {004} reflections (d).

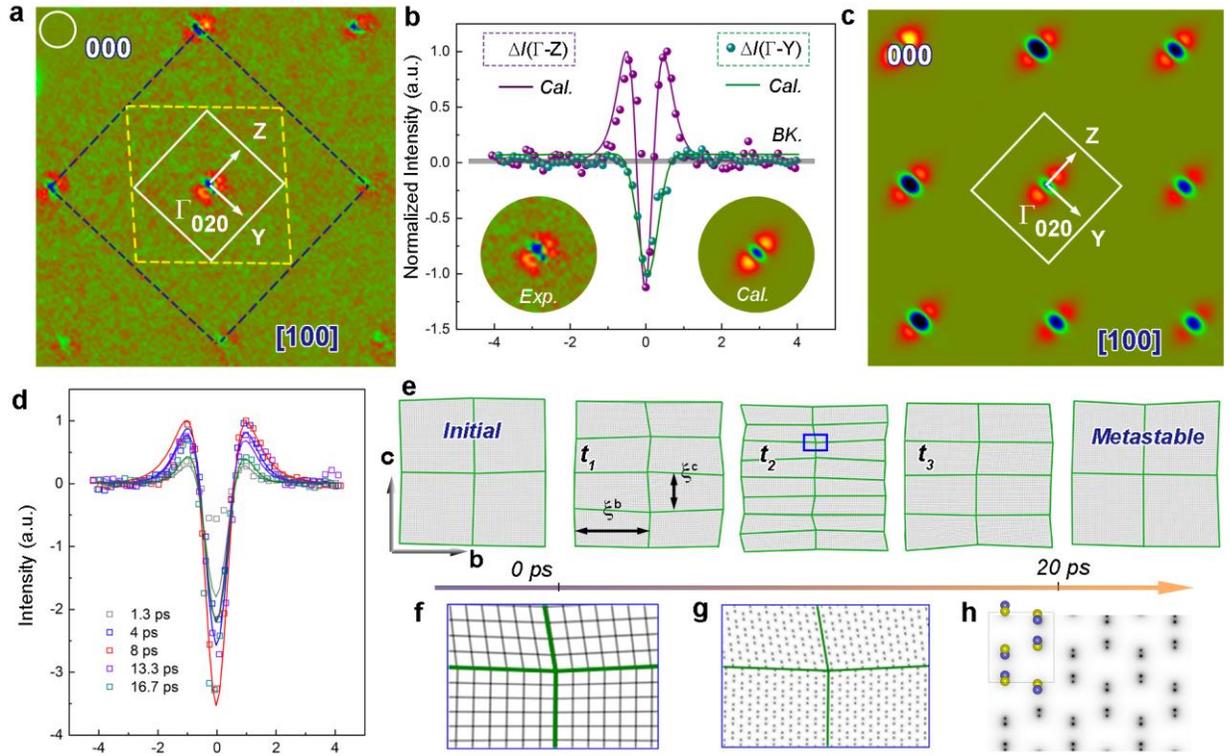

Fig. 2 **Photoinduced domain evolution observed in UED patterns.** (a) An intensity difference map between the UED patterns taken before time-zero and at 1.7 ps. The Brillouin zones around the (020) Bragg peak (Γ point) with the 1st, 2nd and 3rd Brillouin zone boundaries marked by white, yellow and black lines, respectively. The (000) beam position is indicated by the circle. (b) The dots present the line profiles of the diffraction intensity along the Γ-Z (purple dots) and Γ-Y directions (green dots) in (a), showing significant asymmetry of the intensity profiles. The solid lines are the intensity profiles from the simulated pattern in (c). (c) Simulated intensity difference map calculated using Monte Carlo based multislice method with large supercell containing domain structure between $t \leq 0$ (initial) and $t > 0$. (d) Line profiles along Γ-Z direction of the Bragg peak in the UED intensity difference pattern as a function of time. The squares are experimental data, the solid lines are simulation result. (e) Schematic of the domain formation and evolution after photoexcitation, showing the change of domain size ($\xi_c$) along the **c** direction with time, while the domain size ($\xi_b$) along the **b** direction remains unchanged. The domain size in **b** and **c** directions becomes equal in the metastable state. (f) Magnified drawing from the area marked by the blue rectangle in (e), showing lattice distortion after forming domains. (g) Projected potential calculated based on the distorted lattice in (f). The lattice distortion is exaggerated for clarity in (e-g). (h) [100] projected electric potential calculated from the undistorted SnSe structure. The [100] projection of SnSe structure model is embedded with blue and yellow spheres representing Sn and Se, respectively.

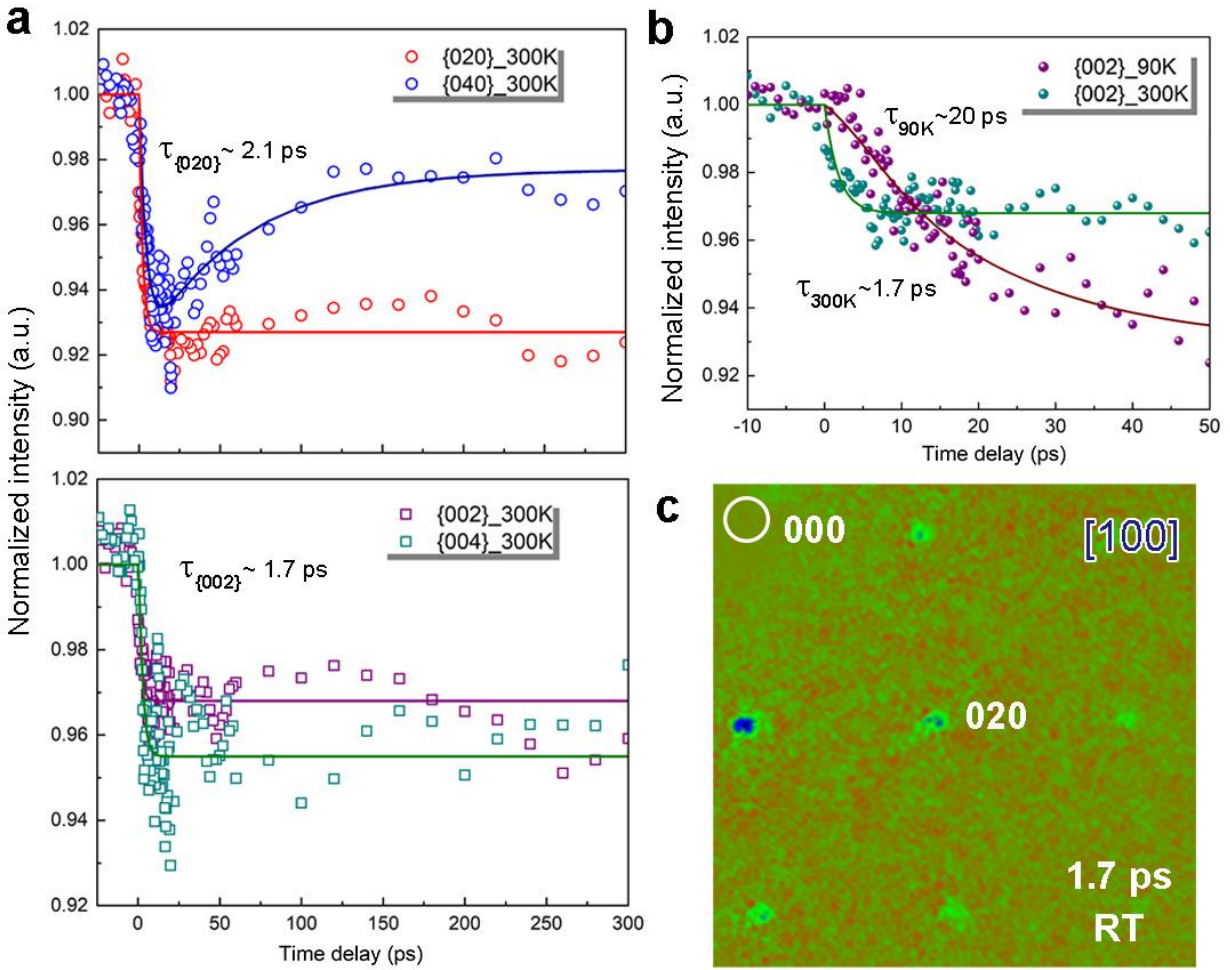

Fig. 3 **UED experimental observation at 300 K.** (a) Temporal evolution of the normalized diffraction intensity $I/I_0$ of Bragg peaks {020}, {040}, {002} and {004} at 90 K. (b) Comparison of the intensity changes of Bragg peaks {002} at 90 K and 300 K. The intensity drops faster at 300 K, i.e., time constant $\tau_{300K} \sim 1.7$ ps $< \tau_{90K} \sim 20$ ps. (c) The isotropic distribution around each Bragg peak in the intensity difference map at 300 K at 1.7 ps after laser pump.

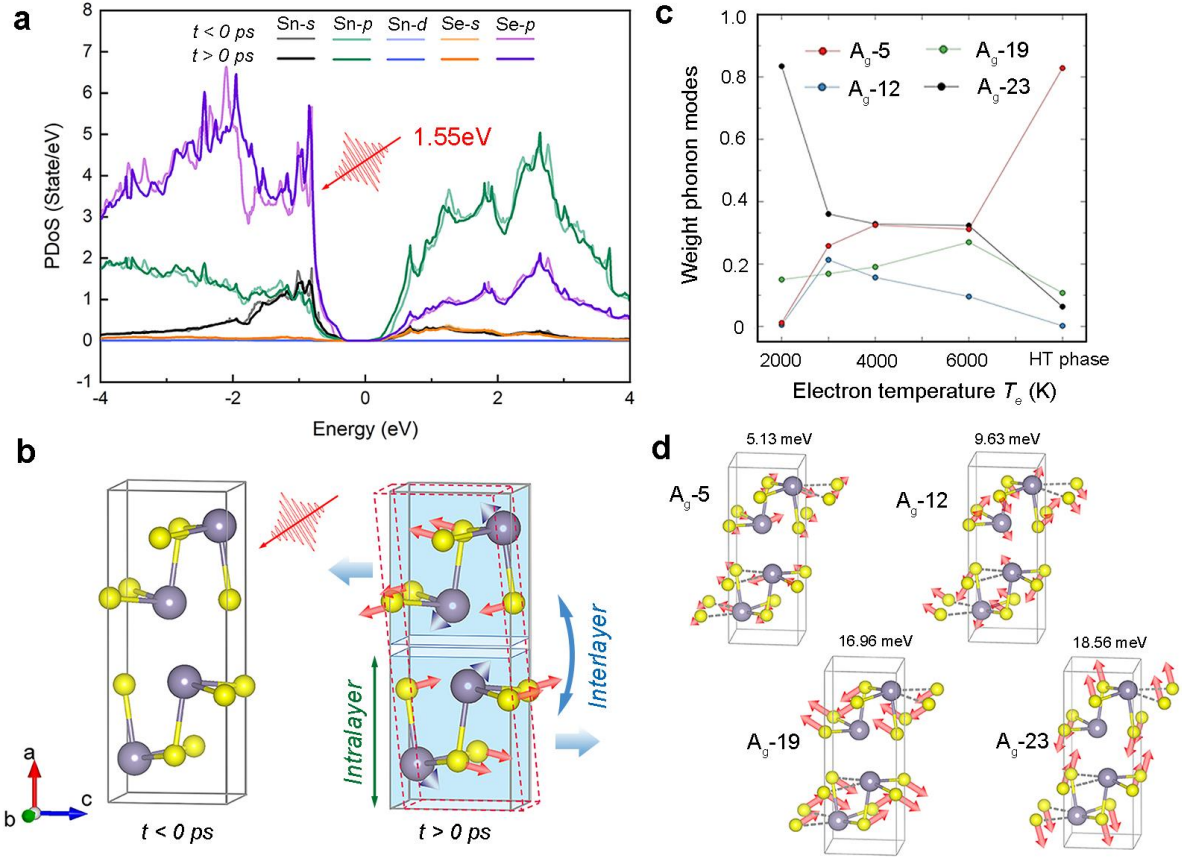

Fig. 4 **DFT calculations.** (a) DFT calculations of partial density of states for the structures relaxed with $T_e = 0$ K (light lines) and $T_e = 6000$ K (thick lines), showing that the 1.55 eV incident photons excite the electrons near the Fermi surface to high energy levels. (b) Photoinduced atomic displacements and relaxation that can be attributed to the transverse optical $A_g$ phonon modes. The Sn-Se movement increases the intralayer lattice symmetry, while induces interlayer shear strain along **c** axis, which yields lattice distortions and formation of domains. (c) Projected weights of the atomic displacements into the four $A_g$ modes of the low-temperature *Pnma* phase for different electron temperature $T_e$. (d) Lattice distortion models corresponding to $A_g$ phonon modes with different energy.

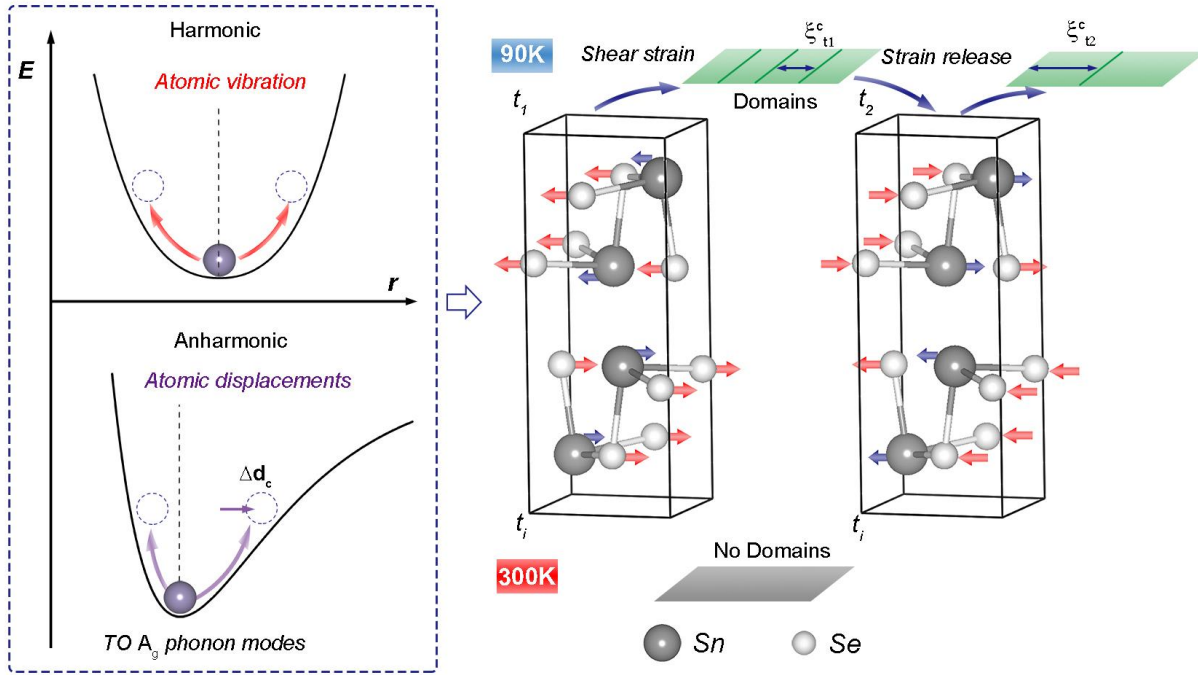

Fig. 5 **Photoinduced domain formation in SnSe.** After photoexcitation, the electrons are 'heated' into excited states. During electron excitation and relaxation processes, via energy flowing from electrons to phonons, in the harmonic energy potential, the atom vibration near the equilibrium position is enhanced, which increases the instability of the lattice periodicity. Additionally, the transverse optical (TO) A$_g$ phonon modes are excited through electron-phonon coupling, atoms deviate from the original positions along **c** axis ($\Delta \mathbf{d_c}$). There are four $A_g$ phonon modes in the **b-c** plane in SnSe. At 90 K, the direction of atomic displacements dominates at different delay times. The incompatibility reduces the correlation length ($\xi_{t1}^c$) generates domains driven by the shear strain between layers. The strain release drives the atoms move to the opposite direction, which gradually increases the correlation length $\xi_{t2}^c$.